\documentclass[ALICE,manyauthors]{cernphprep}
\usepackage[comma,square,numbers,sort&compress]{natbib}
\usepackage{hyperref}
\usepackage{lineno}
\usepackage{xspace}
\usepackage{xcolor}
\usepackage{comment}

\usepackage[T1]{fontenc}
\usepackage{orcidlink} 

\begin{document}
%%%%%%%%%%%%%%%%%%%%%%%%%%%%%%%%%%%%%%%%%%%%%%%%%%
% These are some new commands that may be useful 
% for paper writing in general. If other newcommands
% are needed for your specific paper, please feel 
% free to add here. 
%
% The currently available commands are organized in: 
% 1) Systems
% 2) Quantities
% 3) Energies and units
% 4) Detectors
% 5) particle species 
%%%%%%%%%%%%%%%%%%%%%%%%%%%%%%%%%%%%%%%%%%%%%%%%%%

% 1) SYSTEMS 
\newcommand{\pp}           {pp\xspace}
\newcommand{\ppbar}        {\mbox{$\mathrm {p\overline{p}}$}\xspace}
\newcommand{\XeXe}         {\mbox{Xe--Xe}\xspace}
\newcommand{\PbPb}         {\mbox{Pb--Pb}\xspace}
\newcommand{\pA}           {\mbox{pA}\xspace}
\newcommand{\pPb}          {\mbox{p--Pb}\xspace}
\newcommand{\AuAu}         {\mbox{Au--Au}\xspace}
\newcommand{\dAu}          {\mbox{d--Au}\xspace}

% 2) QUANTITIES 
\newcommand{\s}            {\ensuremath{\sqrt{s}}\xspace}
\newcommand{\snn}          {\ensuremath{\sqrt{s_{\mathrm{NN}}}}\xspace}
\newcommand{\pt}           {\ensuremath{p_{\rm T}}\xspace}
\newcommand{\ptt}           {\ensuremath{p_{\rm T}^{\rm trig}}\xspace}
\newcommand{\pta}           {\ensuremath{p_{\rm T}^{\rm assoc}}\xspace}
\newcommand{\meanpt}       {$\langle p_{\mathrm{T}}\rangle$\xspace}
\newcommand{\ycms}         {\ensuremath{y_{\rm CMS}}\xspace}
\newcommand{\ylab}         {\ensuremath{y_{\rm lab}}\xspace}
\newcommand{\etarange}[1]  {\mbox{$\left | \eta \right |~<~#1$}}
\newcommand{\yrange}[1]    {\mbox{$\left | y \right |~<~#1$}}
\newcommand{\dndy}         {\ensuremath{\mathrm{d}N_\mathrm{ch}/\mathrm{d}y}\xspace}
\newcommand{\dndeta}       {\ensuremath{\mathrm{d}N_\mathrm{ch}/\mathrm{d}\eta}\xspace}
\newcommand{\avdndeta}     {\ensuremath{\langle\dndeta\rangle}\xspace}
\newcommand{\dNdy}         {\ensuremath{\mathrm{d}N_\mathrm{ch}/\mathrm{d}y}\xspace}
\newcommand{\Npart}        {\ensuremath{N_\mathrm{part}}\xspace}
\newcommand{\Ncoll}        {\ensuremath{N_\mathrm{coll}}\xspace}
\newcommand{\dEdx}         {\ensuremath{\textrm{d}E/\textrm{d}x}\xspace}
\newcommand{\RpPb}         {\ensuremath{R_{\rm pPb}}\xspace}

% 3) ENERGIES, UNITS
\newcommand{\nineH}        {$\sqrt{s}~=~0.9$~Te\kern-.1emV\xspace}
\newcommand{\seven}        {$\sqrt{s}~=~7$~Te\kern-.1emV\xspace}
\newcommand{\twoH}         {$\sqrt{s}~=~0.2$~Te\kern-.1emV\xspace}
\newcommand{\twosevensix}  {$\sqrt{s}~=~2.76$~Te\kern-.1emV\xspace}
\newcommand{\five}         {$\sqrt{s}~=~5.02$~Te\kern-.1emV\xspace}
\newcommand{\twosevensixnn}{$\sqrt{s_{\mathrm{NN}}}~=~2.76$~Te\kern-.1emV\xspace}
\newcommand{\fivenn}       {$\sqrt{s_{\mathrm{NN}}}~=~5.02$~Te\kern-.1emV\xspace}
\newcommand{\LT}           {L{\'e}vy-Tsallis\xspace}
\newcommand{\GeVc}         {Ge\kern-.1emV/$c$\xspace}
\newcommand{\MeVc}         {Me\kern-.1emV/$c$\xspace}
\newcommand{\TeV}          {Te\kern-.1emV\xspace}
\newcommand{\GeV}          {Ge\kern-.1emV\xspace}
\newcommand{\MeV}          {Me\kern-.1emV\xspace}
\newcommand{\GeVmass}      {Ge\kern-.1emV/$c^2$\xspace}
\newcommand{\MeVmass}      {Me\kern-.1emV/$c^2$\xspace}
\newcommand{\lumi}         {\ensuremath{\mathcal{L}}\xspace}

% 4) DETECTORS 
\newcommand{\ITS}          {\rm{ITS}\xspace}
\newcommand{\TOF}          {\rm{TOF}\xspace}
\newcommand{\ZDC}          {\rm{ZDC}\xspace}
\newcommand{\ZDCs}         {\rm{ZDCs}\xspace}
\newcommand{\ZNA}          {\rm{ZNA}\xspace}
\newcommand{\ZNC}          {\rm{ZNC}\xspace}
\newcommand{\SPD}          {\rm{SPD}\xspace}
\newcommand{\SDD}          {\rm{SDD}\xspace}
\newcommand{\SSD}          {\rm{SSD}\xspace}
\newcommand{\TPC}          {\rm{TPC}\xspace}
\newcommand{\TRD}          {\rm{TRD}\xspace}
\newcommand{\VZERO}        {\rm{V0}\xspace}
\newcommand{\VZEROA}       {\rm{V0A}\xspace}
\newcommand{\VZEROC}       {\rm{V0C}\xspace}
\newcommand{\Vdecay} 	   {\ensuremath{V^{0}}\xspace}

% 4) PARTICLE SPECIES 
\newcommand{\ee}           {\ensuremath{e^{+}e^{-}}} 
\newcommand{\pip}          {\ensuremath{\pi^{+}}\xspace}
\newcommand{\pim}          {\ensuremath{\pi^{-}}\xspace}
\newcommand{\kap}          {\ensuremath{\rm{K}^{+}}\xspace}
\newcommand{\kam}          {\ensuremath{\rm{K}^{-}}\xspace}
\newcommand{\pbar}         {\ensuremath{\rm\overline{p}}\xspace}
\newcommand{\kzero}        {\ensuremath{{\rm K}^{0}_{\rm{S}}}\xspace}
\newcommand{\lmb}          {\ensuremath{\Lambda}\xspace}
\newcommand{\almb}         {\ensuremath{\overline{\Lambda}}\xspace}
\newcommand{\Om}           {\ensuremath{\Omega^-}\xspace}
\newcommand{\Mo}           {\ensuremath{\overline{\Omega}^+}\xspace}
\newcommand{\X}            {\ensuremath{\Xi^-}\xspace}
\newcommand{\Ix}           {\ensuremath{\overline{\Xi}^+}\xspace}
\newcommand{\Xis}          {\ensuremath{\Xi^{\pm}}\xspace}
\newcommand{\Oms}          {\ensuremath{\Omega^{\pm}}\xspace}
\newcommand{\degree}       {\ensuremath{^{\rm o}}\xspace}

% Our additional commands

\newcommand{\as}{away-side\ }
\newcommand{\ns}{near-side\ }

\newcommand{\As}{Away-side\ }
\newcommand{\Ns}{Near-side\ }

\newcommand{\pTa}{\pt^{\text{assoc}}}

\newcommand{\veve}{\ensuremath{v_{3}^{t}v_{3}^{a}}}
\newcommand{\piz}{\ensuremath{\pi^{0}}}
\newcommand{\pizcand}{\ensuremath{\pi^{0}_{\text{Cand}}}}

%%%%%%%%%%%%%%%  Title page %%%%%%%%%%%%%%%%%%%%%%%%
\begin{titlepage}
% the dates below correspond to CERN approval
% please don't touch: EB chairs will take care
\PHyear{2026}       % required, will be obtained from CERN
\PHnumber{036}      % required, will be obtained from CERN
\PHdate{13 Feb}  % required, will be obtained from CERN
%%%%%%%%%%%%%%%%%%%%%%%%%%%%%%%%%%%%%%%%%%%%%%%%%%%%

%%% Put your own title + short title here:
\title{Measurement of $\piz-$hadron correlations relative to the event plane in semicentral \PbPb collisions at $\mathbf{\sqrt{\textit{s}_{\textup{NN}}} = 5.02~}$TeV}
\ShortTitle{$\piz-$hadron correlations relative to the event plane}   % appears on left page headers

%%% Do not change the next lines
\Collaboration{ALICE Collaboration\thanks{See Appendix~\ref{app:collab} for the list of collaboration members}}
\ShortAuthor{ALICE Collaboration} % appears on right page headers, do not change

% AAAAAAAAAAAAAAAAAAAAAAAAAAAAAAAAAAAAAAAAAAAAAAAAAAAAAAAAAAAAAAAA
% AAAAAAAAAAAAAAAAAAAAAAAAAAAAAAAAAAAAAAAAAAAAAAAAAAAAAAAAAAAAAAAA
% AAAAAAAAAAAAAAAAAAAAAAAAAAAAAAAAAAAAAAAAAAAAAAAAAAAAAAAAAAAAAAAA
\begin{abstract}
% AAAAAAAAAAAAAAAAAAAAAAAAAAAAAAAAAAAAAAAAAAAAAAAAAAAAAAAAAAAAAAAA

Per-trigger yields of $\piz-$hadron correlations were measured in semicentral \PbPb collisions at $\snn$ $=5.02~$\TeV in ALICE at the LHC. The reconstructed $\piz \rightarrow \gamma \gamma$, with a transverse momentum of $11<\pt(\piz)<14~$\GeVc, is used as the trigger particle to calculate yields of associated charged particles on the near- and away-side. The photons are reconstructed using the ALICE Electromagnetic Calorimeter, and the charged particles are measured in the ALICE central barrel within a pseudorapidity range of $|\eta|<0.8$. The yields are reported relative to the orientation of the $\piz$ with the $2^{\rm nd}$-order event plane, and are background subtracted using the reaction plane fit method.
The data give an indication of a suppression of the associated charged-particle yields near $\pt\approx2~$\GeVc when comparing out-of-plane to in-plane trigger particles. At associated charged-particle $\pt > 3$ \GeVc, no significant event-plane dependence is observed within uncertainties. The results are compared with predictions from the JEWEL model, which implements jet energy loss in an expanding medium. JEWEL predicts no significant modification of either the near- or away-side associated yields, independent of whether medium recoils are included. The observed behavior may indicate the presence of additional energy-loss mechanisms beyond those governed by path-length dependence.

\end{abstract}
\end{titlepage}

\setcounter{page}{2} %please do not remove this line

%%%%%%%%%%%%%%%%%%%%%%%%%%%%%%%%
% begin main text
%%%%%%%%%%%%%%%%%%%%%%%%%%%%%%%%
% AAAAAAAAAAAAAAAAAAAAAAAAAAAAAAAAAAAAAAAAAAAAAAAAAAAAAAAAAAAAAAAA
\section{Introduction}
% AAAAAAAAAAAAAAAAAAAAAAAAAAAAAAAAAAAAAAAAAAAAAAAAAAAAAAAAAAAAAAAA

Quantum chromodynamics (QCD), the theory of the strong interaction, predicts a transition from ordinary nuclear matter to a deconfined state of quarks and gluons,  referred to as the quark--gluon plasma (QGP)~\cite{Wilke}. Ultrarelativistic collisions of heavy ions at RHIC and the LHC are able to reach the temperatures necessary to create such a state and enable the study of QGP properties~\cite{HotQCD150, Ratti:2018ksb}. 

During the very early stages of QGP formation, high-energy quarks and gluons are produced from the initial hard scatterings of the colliding nuclei. These partons interact with the medium, fragment, and produce a shower of particles that is referred to as a jet~\cite{Gyulassy:1990ye, JetReviewXinNian, jetreview}. The interactions cause energy loss and angular broadening of partonic showers, which are hypothesized to be proportional to the path length through the medium and fluctuations along the parton's path~\cite{CMS:2012ulu, ATLAS:2018gwx}. As the QGP created in semicentral collisions has significant spatial anisotropies, it is expected that jets lose different amounts of energy depending on their emission angle relative to the $2^{\rm nd}$-order event plane $\Psi_{\rm 2}$~\cite{Wu:2023azi, Ogrodnik:2024qug, ALICE:2010suc, Beattie:2022ojg, ALICE:2023dei} of final-state particles, which approximates the reaction plane defined by the beam axis and the collision impact parameter vector~\cite{ALICE:2010suc, Beattie:2022ojg, ALICE:2023dei}.

The medium-induced modification of jets, or jet quenching, can be studied experimentally by identifying high transverse momentum ($\pt$) hadrons as a proxy for the jet and measuring the yield of associated particles on the near- and away-side of these trigger particles~\cite{STAR:2002svs, PHENIX:2005veb, STARJetHadron, PHENIX:2010hgz, ALICE:2011gpa, STAR:2014wjp, ALICEJetHadron276, STAR:2010uoq, PHENIX:2018wex, STAR:2023hye, PHENIX:2024twd, ALICE:2023qve}. Here, the near-side corresponds to the region in pseudorapidity $\eta$ and azimuthal angle $\varphi$ that is correlated with the direction of the parton shower containing the trigger particle. Conversely, the away-side is the region in $(\eta,\varphi)$ where the high-$\pt$ recoil parton is back-to-back in azimuthal angle to the trigger particle. With sufficient energy loss, a suppression of trigger-associated particle yields is expected at the near- and away-side, with a possible dependence on their orientation relative to the event plane that could indicate a path-length dependence. The quenching of jets has been measured at RHIC and the LHC using inclusive and semi-inclusive hadron and jet production~\cite{ALICEjetRaa, STARjetRaa, CMSjetRaa, ATLASjetRaa}, correlations of dihadrons and dijets~\cite{ATLASdijet, ATLASxj, CMSAj, PHENIX:2009dyg, PHENIX:2012jha, ALICE:2016gso, ALICE:2020atx}, as well as various substructure measurements and modifications of jet fragmentation functions~\cite{ATLAS:2022vii, ALICE:2024jtb}.

One of the challenges of these measurements is the fact that the particle multiplicity in jets is small compared to the number of particles produced by hadronization of the medium. Therefore, the measured trigger--hadron correlations will include a significant combinatorial background, especially towards lower particle $\pt$. Methods to subtract this background, such as the zero-yield-at-minimum (ZYAM) method~\cite{STAR:2005ryu} and the reaction plane fit (RPF) method~\cite{RPFPaper, RPF2, Nattrass:2016cln}, attempt to subtract the momentum anisotropies due to the flow of the background. In the case of the RPF method, this is done by a simultaneous fit to the in-plane, mid-plane, and out-of-plane distributions of the trigger--associated particle pairs in $\Delta \varphi$, where $\Delta \varphi$ is the azimuthal angle between the trigger and associated particle. Then after subtraction, the excess yield of associated particles on the near- and away-side can be obtained by integrating over the $\Delta \varphi$ distribution.

This paper presents the per-trigger yields of $\piz-$hadron correlations in semicentral \PbPb collisions at $\snn=~5.02~$\TeV. The yields are obtained for both the away- and near-side, as well as differentially for the trigger orientation relative to the event plane. The results are compared to model predictions from JEWEL~\cite{JEWELTheory, MilhanoZapp2016, JEWELManual}.

\clearpage

% AAAAAAAAAAAAAAAAAAAAAAAAAAAAAAAAAAAAAAAAAAAAAAAAAAAAAAAAAAAAAAAA
\section{\label{sec:ALICE}Experimental Setup}
% AAAAAAAAAAAAAAAAAAAAAAAAAAAAAAAAAAAAAAAAAAAAAAAAAAAAAAAAAAAAAAAA

The ALICE experiment~\cite{ALICEExperiment} at the LHC is a heavy-ion detector optimized to handle high particle multiplicities over a broad momentum range~\cite{ALICEReview}. The subdetectors used for the analysis presented in this paper are the Inner Tracking System (ITS)~\cite{ALICE:2010tia} and Time Projection Chamber (TPC)~\cite{Alme:2010ke}  for charged-particle reconstruction, the electromagnetic calorimeter (EMCal)~\cite{ALICE:2022qhn} for neutral pions ($\piz \rightarrow \gamma\gamma$), and the V0 detectors~\cite{ALICE:2004ftm} for event-plane determination. For more information on the ALICE apparatus and its performance, see Refs.~\cite{ALICE:2013axi, ALICE:2014sbx, ALICE:2004ftm}

The ITS~\cite{ALICE:2010tia} is a six layer cylindrical silicon detector around the nominal collision point of the experiment. The first two layers are Silicon Pixel Detectors (SPD), followed by two layers of Silicon Drift Detectors (SDD), and two layers of Silicon Strip Detectors (SSD). The TPC~\cite{Alme:2010ke} is a large-volume cylindrical gaseous detector that covers the full azimuthal angle and $|\eta|<0.9$ in pseudorapidity. The information from both the ITS and TPC detectors is used for track reconstruction, covering a transverse-momentum range of $0.15<\pt<30~$\GeVc and a pseudorapidity of $|\eta|<0.9$.

The EMCal~\cite{ALICE:2022qhn} is a sampling calorimeter at a radial distance of 4.28 m from the collision point, with an acceptance of $|\eta| < 0.7$ with $80^{\circ} < \varphi < 187^{\circ}$ plus $0.22< |\eta| < 0.7$ with $260^{\circ} < \varphi < 320^{\circ}$. It provides an energy resolution of $\sigma(E)/E =a \oplus b/\sqrt{E} \oplus c/E$ with parameters $a=1.4 \pm 0.1$, $b=9.5 \pm 0.2~{\mathrm{\GeV}^{1/2}}$, and $c=2.9 \pm 0.9~$\GeV. A clusterization algorithm is used to aggregate the signal of nearby towers into clusters, from which the energy and position are calculated. Photon-like clusters were selected above an energy threshold of $2~$\GeV within $50~$ns of the collision time. Additional background was suppressed by removing those clusters that match with a charged-particle track as reconstructed by the ITS and TPC.

The V0A and V0C~\cite{ALICE:2004ftm} are forward scintillator detectors located at pseudorapidities $2.8~<~\eta~<~5.1$ and $-3.7~<~\eta~<~-1.7$, respectively. 
The sum of the amplitudes measured in the V0 detectors is used to define the centrality classes in \PbPb  collisions. Furthermore, the V0A is used for calculating event-plane angles.

% AAAAAAAAAAAAAAAAAAAAAAAAAAAAAAAAAAAAAAAAAAAAAAAAAAAAAAAAAAAAAAAA
\section{\label{sec:Analysis Methods}Analysis Methods}
% AAAAAAAAAAAAAAAAAAAAAAAAAAAAAAAAAAAAAAAAAAAAAAAAAAAAAAAAAAAAAAAA

The \PbPb data used for the analysis were taken during 2015 and 2018, at a center-of-mass energy per nucleon--nucleon pair of $\snn=5.02~$\TeV. During both runs, the level-1 $\gamma$-trigger of the EMCal was used to trigger on events that contain a high-momentum photon with a minimum energy threshold of $10~$\GeV~\cite{Bourrion_2013}. The analysis uses the data from this trigger as well as from the default minimum-bias trigger, which requires coincident signals in the V0A and V0C. Furthermore, events are required to have 30--50\% collision centrality and a primary vertex within $\pm 10~$cm of the nominal interaction point along the beam line ($z$ direction). Out-of-bunch pile-up was removed by utilizing the correlation between the number of hits in the ITS and TPC detectors. After applying all these event-selection criteria, $1.68$ million EMCal level-1 triggered events as well as $60.8$ million minimum-bias triggered events are used for the analysis.

\subsection{\label{sec:Method:Pi0ID}Neutral pion and charged-particle reconstruction}

\begin{figure}[b]
    \centering
    \includegraphics[width=0.49\linewidth]{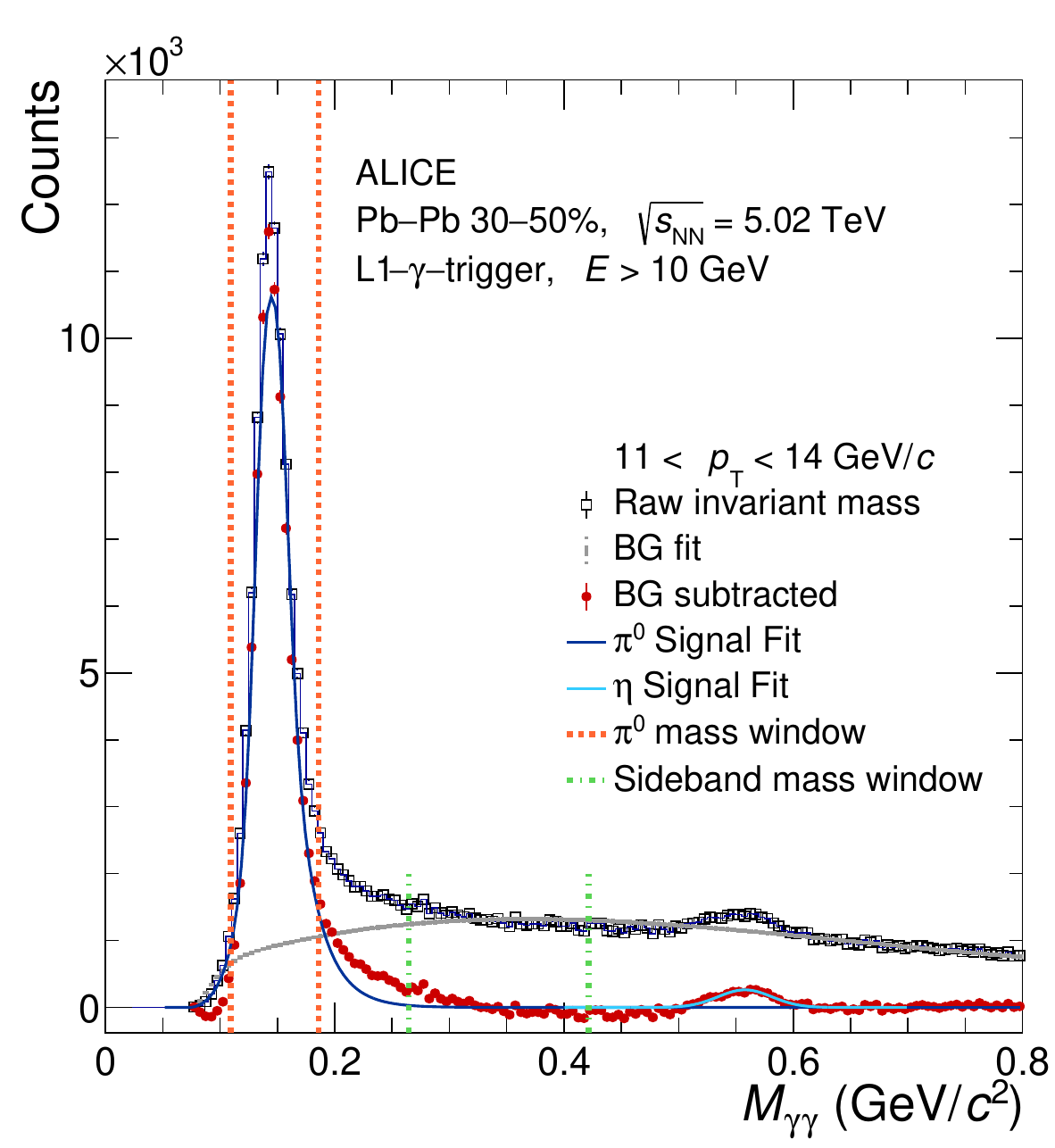}
    \caption{Diphoton invariant mass distribution up to $0.8~$\GeVmass before and after subtraction of the combinatorial background, for pair momenta $11<\pt<14~$\GeVc. The orange and green vertical dashed lines indicate the $\piz$ and selected sideband mass windows, respectively.}
    \label{fig:invarMass}
\end{figure}

Neutral pion $(\piz)$ candidates are reconstructed in the two-photon decay channel $\piz \rightarrow \gamma \gamma$ by calculating the invariant mass of EMCal cluster pairs using $M_{\gamma\gamma} = \sqrt{2 E_{\gamma 1} E_{\gamma 2} (1- \mathrm{cos}(\theta_{\gamma 1 \gamma 2}))}$. Here, $E_{\gamma1}$ and $E_{\gamma2}$ are the energies of the photon-like clusters $1$ and $2$, respectively, and $\theta_{\gamma1\gamma2}$ is the opening angle between the clusters. The minimum energy of the clusters considered is $2~$\GeV, and clusters too close to the border of the calorimeter are rejected, due to possible incorrect energy measurement from losses at the edge. The $\piz$'s are selected with a transverse momentum of $11<\pt<14~$\GeVc. The $\piz$ signal is an excess yield on top of a combinatorial background, see Fig.~\ref{fig:invarMass}. The shape of the peak and combinatorial background are estimated using Gaussian and polynomial fit functions, capturing both the shape of the signal and the background. Note that the $\eta$ meson is visible around $M_{\gamma\gamma}=548~$\GeVmass, which is fitted with a Gaussian but otherwise not used in any part of the analysis other than extending the fit range of the background. The mean ($\mu_{\piz}$) and width ($\sigma_{\piz}$) of the fit are used to define the integration range of the signal, where ($\mu_{\piz}\pm2\sigma_{\piz}$) is the default.

The charged particles, as reconstructed by the ITS and TPC, require at least one hit in the SPD. The yield of charged particles is corrected for finite track reconstruction efficiency using \PbPb events that are simulated using the HIJING Monte Carlo generator in combination with the GEANT3 particle transport code~\cite{Gyulassy:1994ew,Brun:1994aa}. The efficiency is 
parametrized two-dimensionally as a function of $\pt$ and $\eta$ and is on the order of $80\%$ at $|\eta|<0.5$, and decreases to about $70\%$ for larger $|\eta|$.

\subsection{\label{sec:Method:EventPlane} Event-plane determination}

A collision of heavy ions with a non-zero impact parameter creates typically an almond-shaped, or anisotropic, overlap region. The plane spanned by connecting the two centroids of the nuclei with the beam axis is commonly referred to as the reaction plane, and is experimentally not directly accessible. However, we can determine the $n^{\text{th}}$-order event-plane angle using the directions of the final-state particles in the V0 detectors from

\begin{equation}
\Psi_{n} = 
\frac{1}{n}
\arctan\left(\frac{\sum_i w_i\sin(n\varphi_i)}{\sum_i w_i \cos(n\varphi_i)}\right).
    \label{eq:epAngle}
\end{equation}

Here, $\varphi_{i}$ and $w_{i}$ are the azimuthal angle and signal weight, respectively, of the $i$-th segment of the V0A detector~\cite{Voloshin:1994mz,ALICE:2014sbx}. For $n=2$, $\Psi_{2}$,  the $2^{\rm nd}$-order event plane, is the experimental proxy for the reaction plane. The directions parallel and perpendicular to this plane are referred to as in-plane and out-of-plane angles, respectively, which are the angles of greatest interest for this analysis. Furthermore, the direction of the $\piz$ trigger with respect to $\Psi_{2}$ is used to distinguish the cases for which the trigger is in-, mid-, or out-of-plane. Each of these directions is defined by a region of $\pm30$ degrees in $\varphi$ with respect to $\Psi_{2}$, where the mid-plane is between the in- and out-of-plane windows.

To account for the finite resolution of the $\Psi_{n}$~\cite{Poskanzer:1998yz}, we define the experimental resolution of the $m^{\text{th}}$-order event plane for a $v_n = \langle \cos(n(\varphi-\Psi_{n}))\rangle$ flow term as

\begin{equation}
    R_{n,~m}\equiv \langle 
    \cos(n(\Psi_{m}^{\text{meas}}-\Psi_{m}^{\text{truth}}))
    \rangle,
    \label{eq:epr1}
\end{equation}

with the brackets denoting the average over all events. This can be used to associate the measured values of anisotropic flow coefficients $v_n$ to their true values via

\begin{equation}
    v_{n}^{\text{meas}} = R_{n,~m} v_{n}^{\text{truth}}.
    \label{eq:epr2}
\end{equation}

The event-plane resolutions can be determined by measuring the correlation between event-plane angles from different subdetectors, specifically over different ranges in $\eta$. In this analysis, we determine the event-plane resolution of the V0A event-plane angles using the A-side ($\eta > 0$) and C-side ($\eta < 0$) of the TPC. An average value of $R_{n,~m}=0.7712\pm0.0003$ is used to calculate the correction on the ratio of out-of-plane over in-plane yields of $\piz-$hadron correlations.

\subsection{\label{sec:Method:Pi0HadronCorr}\texorpdfstring{$\piz-$}{Pi0}hadron correlations}

The angular correlations between $\piz$ candidates and charged-particle tracks are measured as a function of the $\Delta\eta$ and $\Delta\varphi$ of each pair, as well as the $\pt$ of the associated charged particle and the angle between the $\piz$ trigger and the $2^{\rm nd}$-order event plane. Furthermore, the acceptance of the ALICE detector geometry is corrected using the event-mixing technique. The mixed-event correlations are calculated between $\piz-$hadron candidates, where the $\piz$ and associated charged hadron are explicitly from different events.
The events used for mixing are matched in terms of event-plane angle, centrality, and position of the event vertex. The mixed-event correlation function ($\text{ME}(\Delta\eta,\Delta\varphi)$) is normalized such that its maximum value is unity, and any $\varphi$-dependence in the tracking efficiency cancels in this approach. The correlation between $\piz$ candidates and charged particles from the same event ($\text{SE}(\Delta\eta,\Delta\varphi)$) is divided by those from the mixed events as defined by

\begin{equation}
    C(\Delta\eta,\Delta\varphi) = 
\frac{\text{SE}(\Delta\eta,\Delta\varphi)}{\text{ME}(\Delta\eta,\Delta\varphi)}.
    \label{eq:signal}
\end{equation}

These correlations are projected onto the $\Delta\varphi$ axis in two regions of $\Delta\eta$, namely the near-$\Delta\eta$ ($|\Delta\eta|<0.8$) region that includes the near-side jet peak, and the far-$\Delta\eta$ ($0.8<|\Delta\eta|<1.35$) region where the impact of jet effects is negligible for $|\Delta \varphi|<\pi/2$.

As the $\piz$ candidates have a contamination from combinatorial pairs of EMCal clusters, there is a background contribution to $C(\Delta\eta,\Delta\varphi)$. Therefore, the measured correlation functions of both signal $(C_{\text{sig}})$ and background $(C_{\text{bkg}})$ are related via the purity $P=N_{\text{sig}}/(N_{\text{sig}}+N_{\text{bkg}})$ of the sample via

\begin{equation}
	\label{eq:CorrTotalByPurity}
	\frac{1}{N_{\text{trig}}} C = P \frac{1}{N_{\text{sig}}} C_{\text{sig}} + (1 - P) \frac{1}{N_{\text{bkg}}} C_{\text{bkg}},
\end{equation}

where $N_{\text{trig}}$, $N_{\text{sig}}$, and $N_{\text{bkg}}$, are the number of $\piz$ candidates in the total measured, signal, and background region, respectively. As the background cluster pairs may have different correlations with charged hadrons, this effect must be corrected for an accurate measurement of $\piz-$hadron correlations. To do this, the correlations between background cluster pairs and hadrons are studied using cluster pairs from a sideband region in the diphoton invariant mass, indicated by the green lines in Fig.~\ref{fig:invarMass}, wherein the pairs are assumed to be purely combinatorial. For high-$\pt$ cluster pairs, the lower invariant-mass range is kinematically excluded by the effect of cluster merging, which prevents constructing a sideband on the left side of the $\piz$, as can be seen in Fig.~\ref{fig:invarMass}. To accommodate for this, the correlations with the upper-mass sideband are studied as a function of invariant mass, and extrapolated to the $\piz$-mass region. Having obtained the correlations from the background, the true $\piz-$hadron correlations are extracted using the following equation

\begin{equation}
	\label{eq:CorrSignalSolve}
	\frac{1}{N_{\text{trig}}} \frac{\mathrm{d}^{2}N}{\mathrm{d} \Delta \eta \mathrm{d} \Delta \varphi} \equiv \frac{1}{N_{\text{sig}}} C_{\text{sig}} = \frac{1}{P} \left[\frac{1}{N_{\text{trig}}} C - (1 - P) \frac{1}{N_{\text{bkg}}} C_{\text{bkg}} \right].
\end{equation}

\subsection{\label{sec:Method:BackgroundFlowSub}Subtraction of anisotropic flow}

In heavy-ion collisions, the spatial anisotropies of the initial state are transformed into momentum-space anisotropies of the final-state particles, which are referred to as anisotropic flow. The particle correlations that arise from this are a dominant background when measuring trigger--hadron correlations in $(\Delta \eta, \Delta \varphi)$. In addition, the trigger particle may have correlations with the symmetry planes of the event due to the energy loss being different in- and out-of-plane. When restricting the trigger particles to a specific angular range with respect to the event plane, the subtraction of this background becomes even more important.

The reaction plane fit method (RPF) is used to subtract the contribution from the anisotropic flow of the underlying event~\cite{RPFPaper,RPF2}. This method employs a simultaneous fit of the flow coefficients ${v}_n$ for both the associated particle $({v}_{n}^{~\text{assoc}})$ and the trigger particle $({v}_{n}^{~\text{trig}})$ using

\begin{equation}
    \frac{\mathrm{d}N}{\mathrm{d}\Delta\varphi} = B \left[ 1 + \sum_n 2 {v}_{n}^{~\text{assoc}} {v}_{n}^{~\text{trig}} \cos(n\Delta\varphi) \right],
\end{equation}

where $\Delta \varphi$ is the azimuthal angle between the trigger and associated hadron, $B$ is an overall normalization factor, and $n$ is an integer up to $n=4$. The ${v}_{n}^{~\text{assoc}}$ is expected to be close to the charged-particle $v_{n}$. A similar statement can be made for ${v}_{n}^{~\text{trig}}$, which is consistent with flow measurements at higher transverse momentum. The ${v}_{1}$ parameter is fixed to the results from an independent measurement~\cite{ALICE_V1} to improve the fit quality. For the higher harmonics, instead of fixing the RPF parameters to those of independent measurements, we choose to fit the parameters in a region where no jet-induced trigger--hadron correlations are expected. This region is at far $\Delta \eta$ with $|\Delta \varphi|<\pi/2$, which is then by definition outside the region of the expected jet-like correlations.

\begin{figure}[t]
    \centering
    \includegraphics[width=0.99\linewidth]{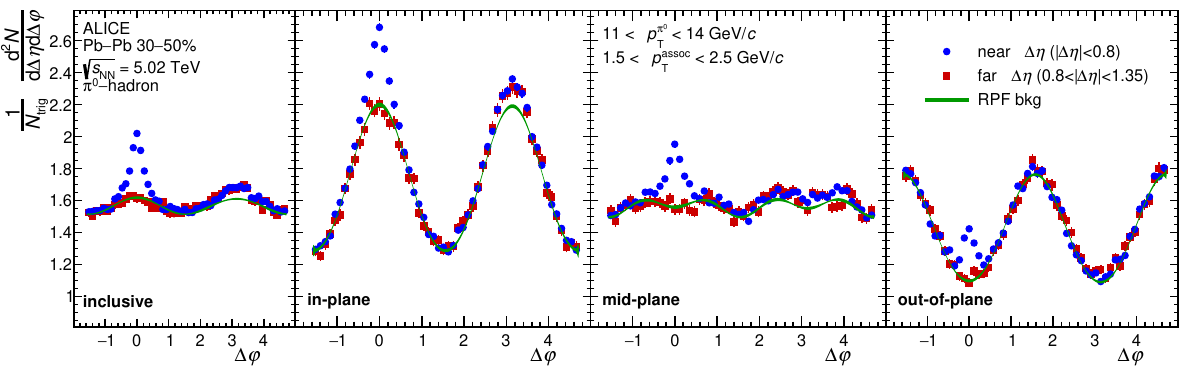}
    \caption{Distributions of $\piz-$hadron correlations as a function of $\Delta \varphi$ in \PbPb collisions at $\snn=5.02~$\TeV, for inclusive, in-, mid-, and out-of-plane orientations of the $\piz-$trigger with respect to the event plane. The near-$\Delta\eta$ ($|\Delta\eta|<0.8$) region that includes the near-side jet peak is shown in blue markers, and the far-$\Delta\eta$ ($0.8<|\Delta\eta|<1.35$) region is shown in red markers. The associated charged hadrons have $1.5<\pta<2.5~$\GeVc. The flow background is fitted using the RPF method and shown by the colored solid lines, where the thickness of the line is the statistical uncertainty of the fit.}
    \label{fig:flowsandwich}
\end{figure}

Figure~\ref{fig:flowsandwich} shows the distributions of $\piz-$hadron correlations as a function of $\Delta \varphi$ in \PbPb collisions at $\snn=5.02~$\TeV, for associated charged hadrons within the transverse-momentum range $1.5<\pta<2.5~$\GeVc. The different panels show the results for inclusive, in-, mid-, and out-of-plane orientations of the $\piz-$trigger with respect to the event plane, with the background fitted using the RPF method. It shows a significant $\Delta \varphi$ modulation that is expected from the magnitude of anisotropic flow coefficients~$v_n$.

\begin{figure}[t]
    \centering
    \includegraphics[width=0.99\linewidth]{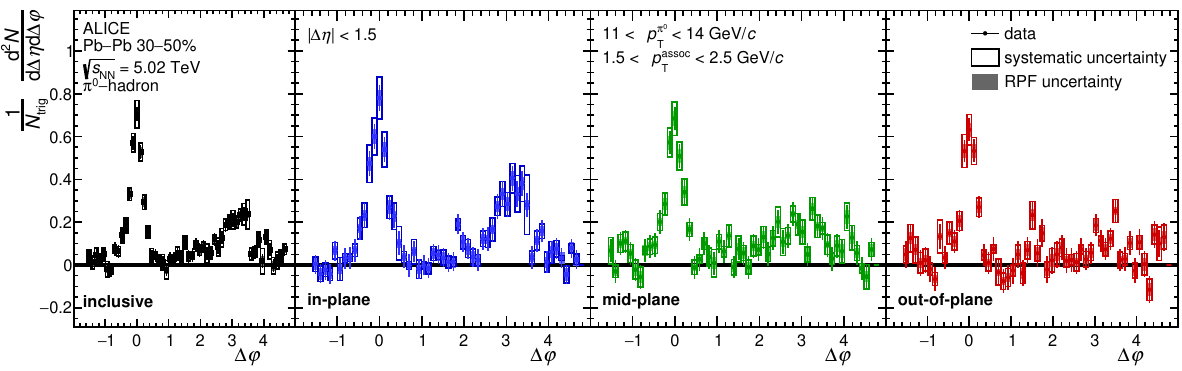}
    \caption{Background-subtracted distributions of $\piz-$hadron correlations as a function of $\Delta \varphi$ in \PbPb collisions at $\snn=5.02~$\TeV for inclusive, in-, mid-, and out-of-plane orientations of the $\piz$ trigger with respect to the event plane. The associated charged hadrons have  $1.5<\pta<2.5~$\GeVc.}
    \label{fig:FlowSub}
\end{figure}

The resulting $\piz-$hadron correlations after subtracting the background are shown in Fig.~\ref{fig:FlowSub}, for associated charged hadrons within the transverse-momentum range $1.5<\pta<2.5~$\GeVc and $|\Delta\eta|<1.5$. For each of the event-plane orientations of the $\piz-$trigger the $\Delta \varphi$ distributions have a significant peak on the near-side, while the peaks on the away-side are less significant, especially for mid-plane and out-of-plane.

\subsection{\label{sec:Method:YieldsAndWidths}Yield of away-side and near-side correlations}
In general, a direct hard-scattering component of the $\piz-$hadron correlations appears as a back-to-back signal in the azimuthal angle $\varphi$. We therefore expect this to manifest as an excess of the yield of associated charged hadrons on the away-side and near-side, relative to the direction of the $\piz$ trigger. These peaks are studied quantitatively by calculating their yields and widths. The yield is calculated using a double integral in $\Delta \eta$ and $\Delta \varphi$ as

\begin{equation}
    Y =
    \int_{\Delta \eta}
    \int_{\Delta \varphi}
    \frac{1}{N_{\text{trig}}} \frac{\mathrm{d}^{2}N}{\mathrm{d}(\Delta \eta) \mathrm{d}(\Delta \varphi)}
    \mathrm{d}(\Delta \varphi) \mathrm{d}(\Delta \eta).
    \label{eq:yield}
\end{equation}

For the near-side peak, the $\Delta \varphi$ limits are $-\pi/3 < \Delta \varphi < \pi/3$ rad. For the away-side peak, the limits are $2\pi/3 < \Delta \varphi <4\pi/3$ rad.
Since the far-$\Delta\eta$ region on the near-side is used to calculate the background with the RPF, that region is not used in order to avoid auto-correlation effects. Therefore the $\Delta \eta$ limits for the near-side are given by $|\Delta \eta|<0.8$, and on the away-side by $|\Delta \eta|<1.35$. 

Furthermore, uncertainties of the RPF parameters introduce uncertainties on the extracted yield via changes in the background subtraction. This uncertainty on the yield is labeled the RPF uncertainty and is calculated by varying the RPF parameters according to their covariance matrix. The RPF uncertainty is reported as a separate statistical uncertainty on the yield of $\piz-$hadron correlations.

\clearpage

% AAAAAAAAAAAAAAAAAAAAAAAAAAAAAAAAAAAAAAAAAAAAAAAAAAAAAAAAAAAAAAAA
\section{\label{sec:Systematics}Systematic Uncertainties}
% AAAAAAAAAAAAAAAAAAAAAAAAAAAAAAAAAAAAAAAAAAAAAAAAAAAAAAAAAAAAAAAA

The systematic uncertainties related to the extracted yield of $\piz-$hadron correlations on the near-side and away-side are listed in Table~\ref{tab:systematics:yields} for the inclusive, in-plane, mid-plane, and out-of-plane trigger orientations. Table~\ref{tab:systematics:ratios} shows the systematic uncertainties for the yield ratios of out-of-plane to in-plane, which are calculated directly on the ratio taking into account the correlations and are found to partially cancel. The systematic uncertainties are indicated by a range that corresponds to the minimum and maximum size of the respective uncertainty as a function of $\pta$. The total systematic uncertainty of the measurement is calculated by treating the separate contributions as uncorrelated and adding them in quadrature. The sources of systematic uncertainties will now be discussed briefly.

\textbf{Purity of the $\piz$-trigger.} As the $\piz$ candidates are reconstructed from two photon-like clusters, they contain combinatorial background from random pairings. The systematic uncertainty of the purity is obtained by varying the fit functions and fit range of the $M_{\gamma \gamma}$ distribution of Fig.~\ref{fig:invarMass}. The yields on the away-side and near-side have been recalculated using the variations in purity, of which the RMS is taken as the systematic uncertainty. The relative systematic uncertainty assigned to this source is at most $7\%$.

\textbf{$\piz$-trigger sideband subtraction.} The correlation $C_{\rm bkg}$ is estimated by changing the sidebands in the $M_{\gamma\gamma}$ invariant-mass distribution, shown as green dot-dashed lines in Fig.~\ref{fig:invarMass}. This sideband has been varied, as well as the methods to extrapolate to the $\piz$-mass region. This relative uncertainty ranges from 1--27$\%$, depending on the event-plane orientation and $\pta$, and is overall slightly higher on the near-side.

\textbf{$\Delta \eta$ and $\Delta \varphi$ selection.} The range in $\Delta \eta$ that defines the near-side is by default chosen to be $|\Delta \eta|<0.8$ and for the away-side $|\Delta \eta|<1.35$. Since the yields are normalized to $\Delta \eta$, any variation due to this choice is corrected for. However, depending on the range, it is possible to either include or exclude regions of true $\piz-$hadron correlations. As such, $\Delta \eta$ and $\Delta \varphi$ have been varied, and the effect on the associated charged-hadron yield has been calculated. At most, these relative uncertainties are $27\%$ for $\Delta \eta$ and $18\%$ for $\Delta \varphi$.

\textbf{RPF method.} The fit parameters of the RPF method, that vary as a function of $\pta$, have uncertainties that directly propagate to an uncertainty in the $\piz-$hadron correlations, and have been accounted for in the statistical uncertainty of the measurement. However, the choice of fixing the fit parameters using input from independent measurements is a source of systematic uncertainty. The variations include fixing the ${v}_{2}^{\rm ~trig}$ parameters to values found with the RPF at lower ${v}_{2}^{\rm ~assoc}$, where the ${v}_{2}^{\rm ~trig}$ is the same, but with better statistical precision. The ${v}_{1}$ parameter is estimated from independent measurements~\cite{ALICE_V1}, whose uncertainty is propagated here. The assigned relative systematic uncertainty for the ${v}_{2}^{\rm ~trig}$ treatment is at most $17\%$ for out-of-plane yields, and the relative ${v}_{1}^{\rm ~trig}{v}_{1}^{\rm ~assoc}$ uncertainty is negligible $(<1\%)$.

\textbf{Charged-particle tracking efficiency.} The relative uncertainty on the tracking efficiency of charged-particle tracks amounts to $4\%$, and results in a scale uncertainty on the yields of $\piz-$hadron correlations. This uncertainty has been obtained by varying the tracking parameters that are used for charged tracks matched between the ITS and TPC. In addition, the event-plane dependence of the tracking efficiency has been studied and found to be negligible.

\textbf{Event-plane resolution.} The event-plane resolution varies slightly with centrality, on the order of $\pm10\%$ from the mean centrality $(40\%)$ to more central and peripheral collisions. A variation on $R_{n,~m}$ of  $4\%$ was chosen, and the effect on the $\piz-$hadron yields was used to calculate the corresponding relative systematic uncertainty. It was found to be at most $4\%$ at low \pta on the near-side, and $1\%$ for higher \pta on the away-side.

\begin{table}
\centering
\caption{Sources of relative systematic uncertainties for the yield of near-side $(|\Delta\eta|<0.8)$ and away-side $(|\Delta\eta|<1.35)$ $\piz-$hadron correlations in \PbPb collisions at $\snn=5.02~$\TeV, for the inclusive, in-, mid-, and out-of-plane orientations of the $\piz$-trigger with respect to the event plane.}
\scalebox{0.75}{
\begin{tabular}{l|cccc|cccc}
     & \multicolumn{4}{c|}{\textbf{near-side}} & \multicolumn{4}{c}{\textbf{away-side}} \\
     & inclusive & in-plane & mid-plane & out-of-plane
     & inclusive & in-plane & mid-plane & out-of-plane\\ [0.9ex]
\hline\hline
    $\piz$ purity         & $1-5\%$ & $1-7\%$  & $1-6\%$ & $1-5\%$ 
        & $0-5\%$ & $0.5-6\%$ & $0.5-5\%$ & $0-3\%$ \\[1ex] 
    
    Sideband subtraction  & $3-18\%$ & $2-14\%$  & $3-27\%$ & $2-17\%$    
        & $1-16\%$ & $3-8\%$ & $2-26\%$ & $2-10\%$  \\[1ex] 
    
    $\Delta\eta$ selection     & $7-20\%$ & $13-25\%$  & $14-23\%$ & $13-24\%$    
        & $5-14\%$ & $13-22\%$ & $15-27\%$ & $17-21\%$         \\[1ex] 

    $\Delta\varphi$ selection  & $0.5-4\%$ & $0.5-3\%$  & $0.5-6\%$ & $0.5-3\%$    
        & $3-17\%$ & $3-17\%$ & $3-18\%$ & $4-14\%$         \\[1ex] 

    ${v}_{2}^{\rm ~trig}$ treatment & $0-1\%$ & $0-7\%$  & $0-6\%$ & $0-17\%$ 
        & $0-1\%$ & $0-10\%$ & $0-8\%$ & $0-18\%$         \\[1ex] 
    
    ${v}_{1}^{\rm ~trig}{v}_{1}^{\rm ~assoc}$ calculation    & $<1\%$ & $<1\%$  & $<1\%$ & $<1\%$ 
        & $<1\%$ & $<1\%$ & $<1\%$ & $<1\%$        \\[1ex] 

    Tracking efficiency        & $4\%$ & $4\%$  & $4\%$ & $4\%$ 
        & $4\%$ & $4\%$ & $4\%$ & $4\%$        \\[1ex] 
    
\hline
    \textbf{Total}              & $13-25\%$ & $18-27\%$  & $19-34\%$ & $15-31\%$    & $9-23\%$ & $18-26\%$ & $18-42\%$ & $15-28\%$       \\[1ex]
\hline
\end{tabular}
}
\label{tab:systematics:yields}
\end{table}

\begin{table}
\centering
\caption{Sources of relative systematic uncertainties for the ratio of out-of-plane over in-plane yields of $\piz-$hadron correlations on the near-side $(|\Delta\eta|<0.8)$ and away-side $(|\Delta\eta|<1.35)$.}
\scalebox{0.75}{
\begin{tabular}{l|cc|cc}
     & \multicolumn{2}{c|}{\textbf{near-side}} & \multicolumn{2}{c}{\textbf{away-side}} \\
     & $\pta < 4$ \GeV/$c$ & $\pta > 4$ \GeV/$c$
     & $\pta < 4$ \GeV/$c$ & $\pta > 4$ \GeV/$c$\\
\hline\hline
    $\piz$ purity           & $1-3\%$  & $1-3\%$   & $1-5\%$ & $1-4\%$  \\[1ex] 
    Sideband subtraction    & $2-11\%$ & $1-3\%$   & $4-17\%$  & $3-12\%$   \\[1ex] 
    $\Delta\eta$ selection       & $12-17\%$& $1-13\%$  & $10-50\%$  & $6-11\%$   \\[1ex] 
    $\Delta\varphi$ selection    & $4-6\%$  & $1\%$     & $6-10\%$  & $1-4\%$   \\[1ex] 
    ${v}_{2}^{\rm ~trig}$ treatment   & $1-11\%$ & $<1\%$    & $4-33\%$  & $1\%$   \\[1ex] 
    ${v}_{1}^{\rm ~trig}{v}_{1}^{\rm ~assoc}$ calculation   & $<1\%$  & $<1\%$& $<1\%$  & $<1\%$   \\[1ex] 
    Tracking efficiency     & $<1\%$   & $<1\%$    & $<1\%$  & $<1\%$   \\[1ex] 
    Event-plane res. corr.  & $1-4\%$  & $1\%$     & $1\%$  & $1\%$   \\[1ex] 

\hline
    \textbf{Total}              & $13-35\%$ & $ 3-14\%$     & $26-65\%$ & $6-16\%$      \\[1ex]
\hline
\end{tabular}
}
\label{tab:systematics:ratios}
\end{table}

\clearpage

% AAAAAAAAAAAAAAAAAAAAAAAAAAAAAAAAAAAAAAAAAAAAAAAAAAAAAAAAAAAAAAAA
\section{\label{sec:Results}Results}
% AAAAAAAAAAAAAAAAAAAAAAAAAAAAAAAAAAAAAAAAAAAAAAAAAAAAAAAAAAAAAAAA

\begin{figure}[htb]
    \centering
    \includegraphics[width=0.49\linewidth]{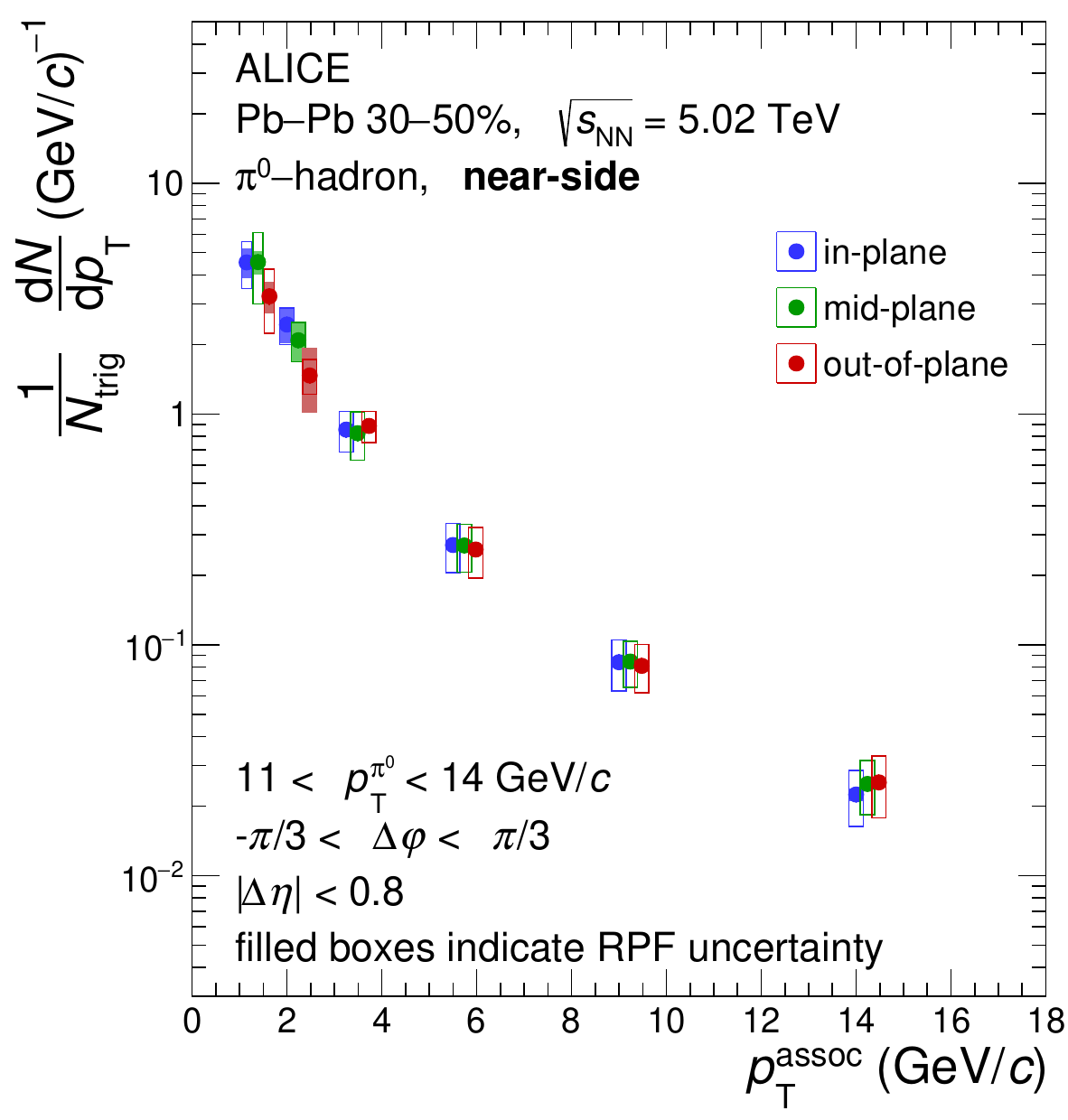}
    \includegraphics[width=0.49\linewidth]{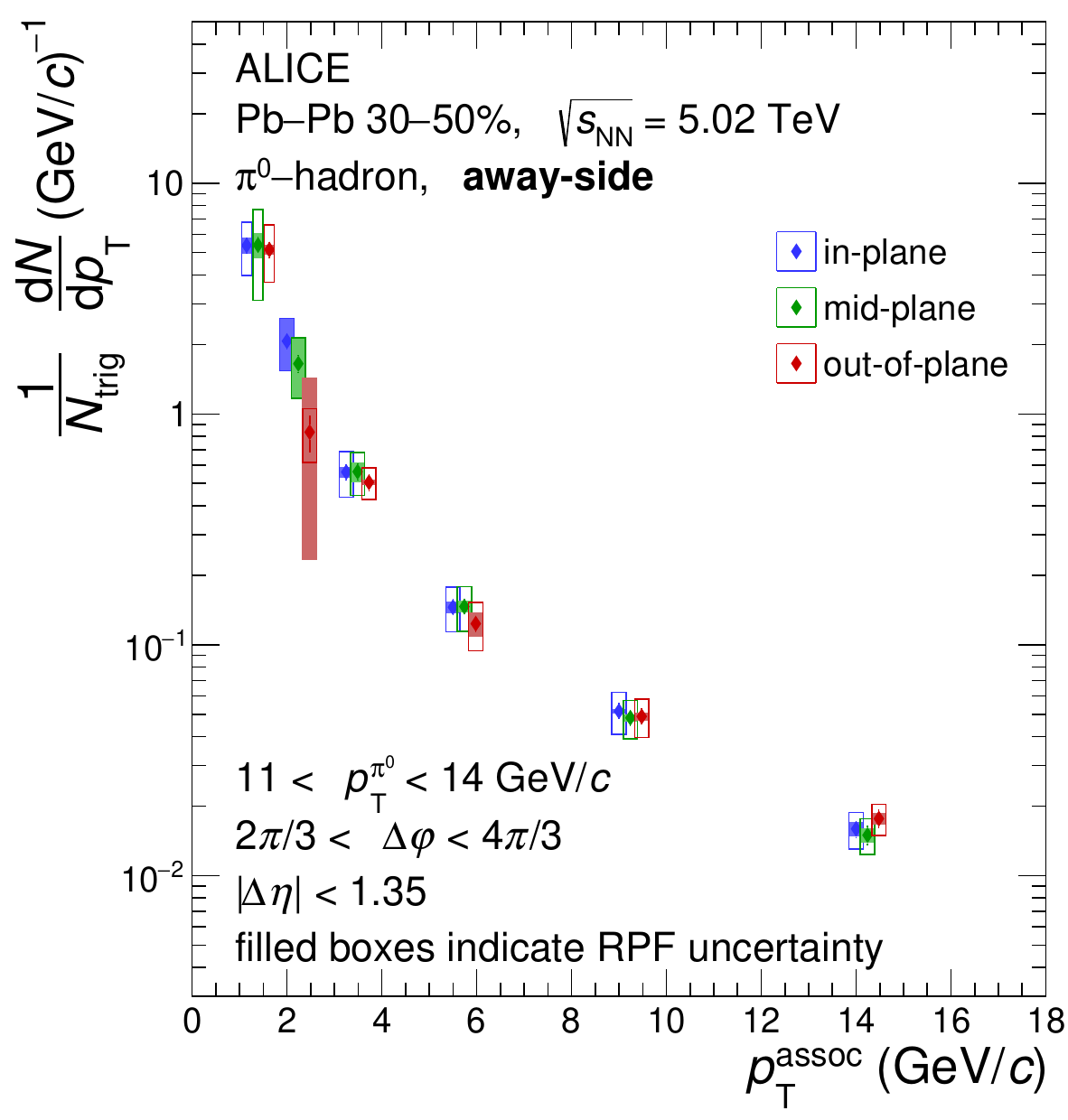}
    \caption{Yield of near-side $(|\Delta\eta|<0.8)$ and away-side $(|\Delta\eta|<1.35)$ $\piz-$hadron correlations as a function of $\pta$ in \PbPb collisions at $\snn=5.02~$\TeV, for the in-, mid-, and out-of-plane orientations of the $\piz$-trigger with respect to the event plane. The data points are slightly displaced for visibility.}
    \label{fig:RawYields}
\end{figure}

The yields of $\piz-$hadron correlations on the near-side and away-side in \PbPb collisions at $\snn=5.02~$\TeV are shown in Fig.~\ref{fig:RawYields} as a function of $\pta$ for in-, mid-, and out-of-plane orientations of the $\piz-$trigger. The $\piz$ candidates were reconstructed with a transverse momentum within the range $11<\pt<14~$\GeVc. For the near-side, the difference in pseudorapidity of the pairs is $|\Delta\eta|<0.8$, while for the away-side, it is $|\Delta\eta|<1.35$. The RPF method is used to subtract the anisotropic flow background for each interval in $\pta$. The results for both near- and away-side follow a steeply falling spectrum, with mostly insignificant changes in the yield for the different event plane distinctions at a given $\pta$. Note that the RPF uncertainty for $\pta = 1.5-2.5$~\GeVc on both the near and away side is much larger than for any of the other bins. Here, the variations of the RPF parameters, specifically the flow coefficients $({v}_{n}^{~\text{assoc}})$, introduce a large variation on the extracted yield.

\begin{figure}
    \centering
    \includegraphics[width=0.49\linewidth]{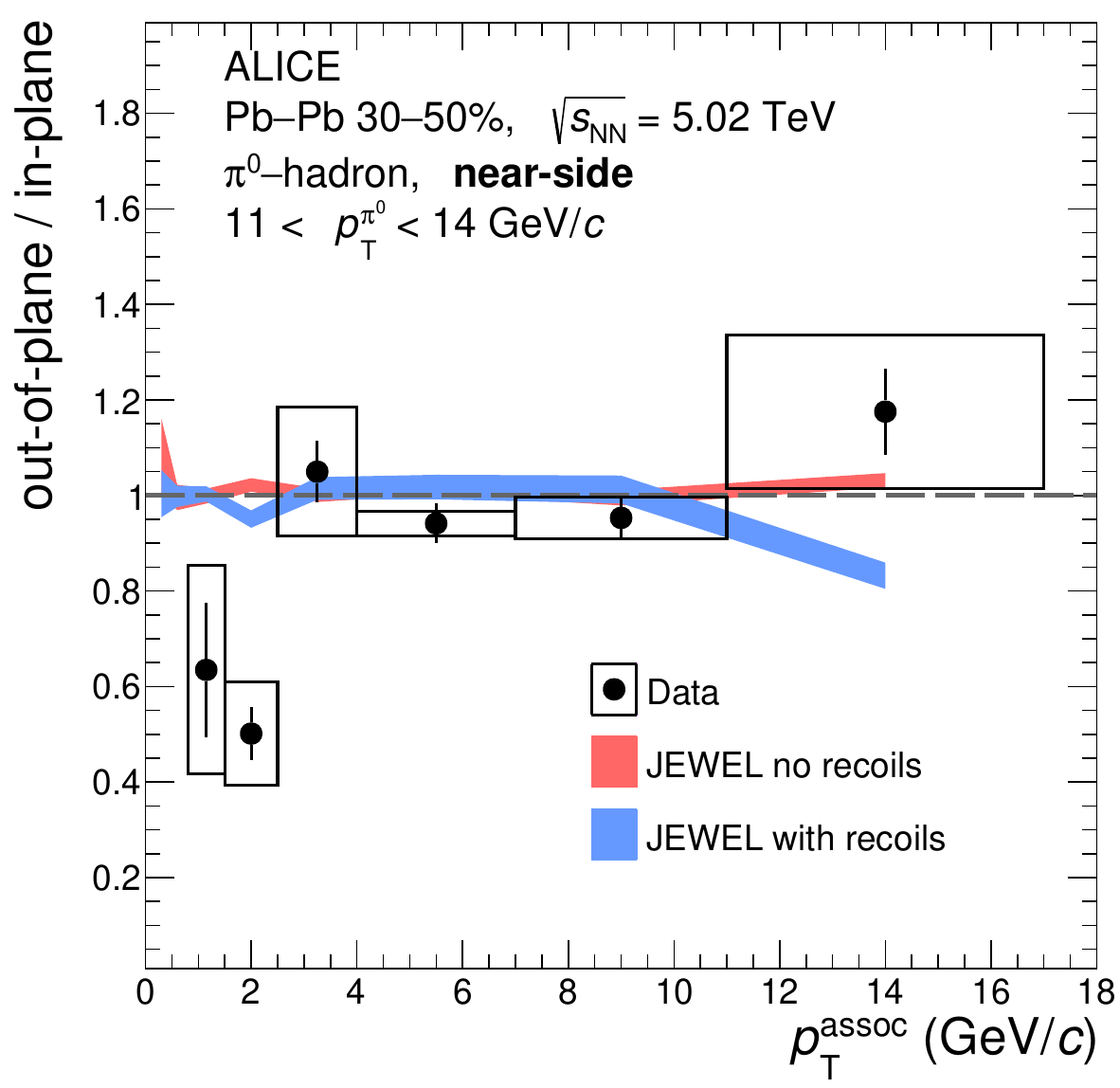}
    \includegraphics[width=0.49\linewidth]{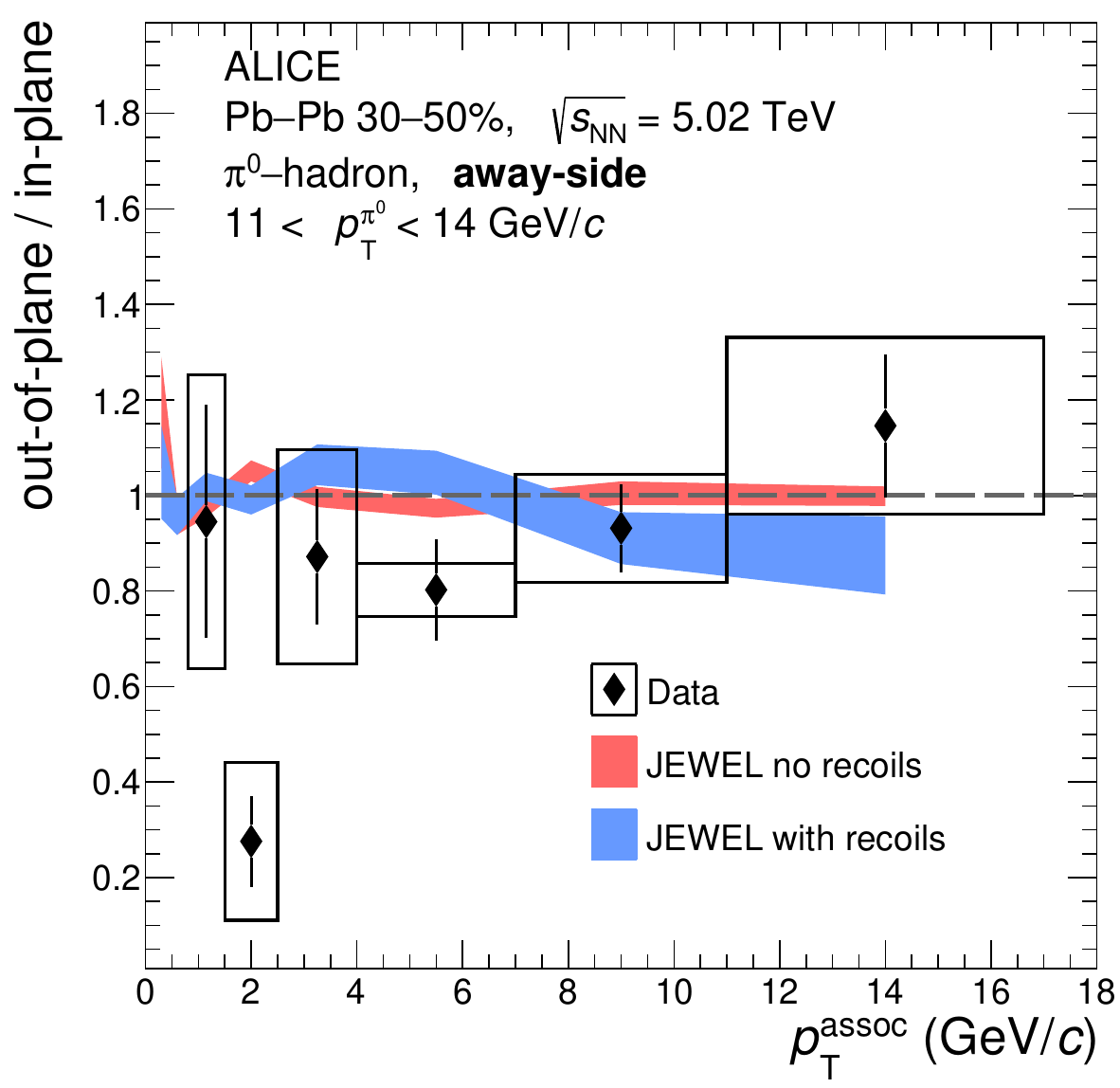}
    \caption{Ratio of out-of-plane over in-plane yields of $\piz-$hadron correlations on the near-side $(|\Delta\eta|<0.8)$ and away-side $(|\Delta\eta|<1.35)$, as a function of $\pta$ in \PbPb collisions at $\snn=5.02~$\TeV.}
    \label{fig:OutOverInWithModels}
\end{figure}

The event-plane dependence is studied by comparing the ratio of the yields for out-of-plane trigger orientations to in-plane trigger orientations. This is done for the near-side and the away-side peaks, with the results shown in Fig.~\ref{fig:OutOverInWithModels}. Note that the event-plane resolution correction, as calculated by Eq.~(\ref{eq:epr2}), is applied on these ratios, and systematic uncertainties partially cancel in the ratio. A deviation from unity might indicate an event-plane dependence of trigger--hadron measurements. In particular, suppression at higher $\pta$ could indicate event-plane-dependent jet-energy loss, especially on the away-side. Such energy loss could also produce a ratio above unity at lower $\pta$ due to energy conservation, as seen in the analysis of jet--hadron correlations in \AuAu and \pp collisions in Ref.~\cite{STARJetHadron}. The results are generally consistent with unity, except at $\pta = 1.5-2.5$~\GeVc for both the near-side and away-side peaks, where the ratios are below unity. This possibly indicates path-length dependent energy loss that has a larger impact on  charged particles at lower $\pta$, as, on average, particles produced out-of-plane traverse a longer path-length in the medium than particles produced in-plane. These observations are consistent with Ref.~\cite{ALICEJetHadron276}.

In addition, the results are compared to predictions made using the JEWEL 2.2.0 model, for the interaction of jets with the QGP medium~\cite{MilhanoZapp2016, JEWELManual}. The JEWEL model was run to match the experimental conditions and produce the same observable, using the two main modes of JEWEL, which are referred to as JEWEL without recoils and JEWEL with recoils. The difference between the modes is whether or not the recoiling partons from the jet's interaction with the medium are saved into the event to participate in hadronization with the jet. JEWEL with recoils should properly simulate the conservation of the jet's lost energy, but introduces a non-trivial background not present in JEWEL without recoils. For both the near-side and the away-side, JEWEL predicts no significant deviation from unity, consistent with the measurement within the uncertainties. However, the data exhibit a stronger suppression at $\pta = 1.5-2.5$~\GeVc.

\begin{comment}
\begin{figure}[htb]
    \centering
    \includegraphics[width=0.49\linewidth]{figures/Results/FinalOutOverIn_ASWithModels_UnZoom.pdf}
    \includegraphics[width=0.49\linewidth]{figures/Results/FinalOutOverIn_NSWithModels_UnZoom.pdf}
    \caption{Out/In Yield Ratios for the away-side and near-side, with comparison to JEWEL w/o recoils.}
    \label{fig:OutOverInNSWithModels}
\end{figure}
\end{comment}

% AAAAAAAAAAAAAAAAAAAAAAAAAAAAAAAAAAAAAAAAAAAAAAAAAAAAAAAAAAAAAAAA
\section{\label{sec:Conclusion}Conclusions}
% AAAAAAAAAAAAAAAAAAAAAAAAAAAAAAAAAAAAAAAAAAAAAAAAAAAAAAAAAAAAAAAA

The per-trigger yields of $\piz$-hadron correlations are presented in this paper for semicentral \PbPb collisions at $\snn=~5.02~$\TeV. The $\piz-$trigger was reconstructed using its decay into two photons and with a transverse momentum of $11<\pt<14~$\GeVc. The yields are reported relative to the orientation of the $\piz$-trigger with the second-order event plane, and are background subtracted using the reaction plane fit method.

The out-of-plane and in-plane yields are fully consistent within $1\sigma$ for $\pta>2.5$ \GeVc, on both the near-side and the away-side. At lower $\pta$, the correlations exhibit a deviation from unity that indicates an event-plane dependent suppression. However, JEWEL predicts minimal event-plane dependence and does not deviate from unity by more than a few percent. This might indicate there are other significant sources of energy loss besides the component related to its path length. The effects at lower momenta should be explored by using model calculations that propagate a full parton shower in an event-by-event hydrodynamical simulation. Additionally, future measurements using a prompt photon as the trigger particle are expected to be less susceptible to jet-selection biases and could therefore exhibit a stronger event-plane dependence. Such measurements would provide a more precise comparison with increasingly comprehensive theoretical models, thereby offering improved sensitivity to the path-length dependence of parton energy loss.

%%%%%%%%%%%%%%%%%%%%%%%%%%%%%%%%
% end main text 
%%%%%%%%%%%%%%%%%%%%%%%%%%%%%%%%

%%%%% acknowledgements - handled by EB chairs 
\newenvironment{acknowledgement}{\relax}{\relax}
\begin{acknowledgement}
\section*{Acknowledgements}
% add specific acknowledgements here 
% ...but please don't remove the line below: funding agencies
% will be acknowledged with a custom tex file handled by EB chairs after Collab Round 2
% Version: 2026-02-03

The ALICE Collaboration would like to thank all its engineers and technicians for their invaluable contributions to the construction of the experiment and the CERN accelerator teams for the outstanding performance of the LHC complex.
The ALICE Collaboration gratefully acknowledges the resources and support provided by all Grid centres and the Worldwide LHC Computing Grid (WLCG) collaboration.
The ALICE Collaboration acknowledges the following funding agencies for their support in building and running the ALICE detector:
A. I. Alikhanyan National Science Laboratory (Yerevan Physics Institute) Foundation (ANSL), State Committee of Science and World Federation of Scientists (WFS), Armenia;
Austrian Academy of Sciences, Austrian Science Fund (FWF): [M 2467-N36] and Nationalstiftung f\"{u}r Forschung, Technologie und Entwicklung, Austria;
Ministry of Communications and High Technologies, National Nuclear Research Center, Azerbaijan;
Rede Nacional de Física de Altas Energias (Renafae), Financiadora de Estudos e Projetos (Finep), Funda\c{c}\~{a}o de Amparo \`{a} Pesquisa do Estado de S\~{a}o Paulo (FAPESP) and The Sao Paulo Research Foundation  (FAPESP), Brazil;
Bulgarian Ministry of Education and Science, within the National Roadmap for Research Infrastructures 2020-2027 (object CERN), Bulgaria;
Ministry of Education of China (MOEC) , Ministry of Science \& Technology of China (MSTC) and National Natural Science Foundation of China (NSFC), China;
Ministry of Science and Education and Croatian Science Foundation, Croatia;
Centro de Aplicaciones Tecnol\'{o}gicas y Desarrollo Nuclear (CEADEN), Cubaenerg\'{\i}a, Cuba;
Ministry of Education, Youth and Sports of the Czech Republic, Czech Republic;
The Danish Council for Independent Research | Natural Sciences, the VILLUM FONDEN and Danish National Research Foundation (DNRF), Denmark;
Helsinki Institute of Physics (HIP), Finland;
Commissariat \`{a} l'Energie Atomique (CEA) and Institut National de Physique Nucl\'{e}aire et de Physique des Particules (IN2P3) and Centre National de la Recherche Scientifique (CNRS), France;
Bundesministerium f\"{u}r Forschung, Technologie und Raumfahrt (BMFTR) and GSI Helmholtzzentrum f\"{u}r Schwerionenforschung GmbH, Germany;
National Research, Development and Innovation Office, Hungary;
Department of Atomic Energy Government of India (DAE), Department of Science and Technology, Government of India (DST), University Grants Commission, Government of India (UGC) and Council of Scientific and Industrial Research (CSIR), India;
National Research and Innovation Agency - BRIN, Indonesia;
Istituto Nazionale di Fisica Nucleare (INFN), Italy;
Japanese Ministry of Education, Culture, Sports, Science and Technology (MEXT) and Japan Society for the Promotion of Science (JSPS) KAKENHI, Japan;
Consejo Nacional de Ciencia (CONACYT) y Tecnolog\'{i}a, through Fondo de Cooperaci\'{o}n Internacional en Ciencia y Tecnolog\'{i}a (FONCICYT) and Direcci\'{o}n General de Asuntos del Personal Academico (DGAPA), Mexico;
Nederlandse Organisatie voor Wetenschappelijk Onderzoek (NWO), Netherlands;
The Research Council of Norway, Norway;
Pontificia Universidad Cat\'{o}lica del Per\'{u}, Peru;
Ministry of Science and Higher Education, National Science Centre and WUT ID-UB, Poland;
Korea Institute of Science and Technology Information and National Research Foundation of Korea (NRF), Republic of Korea;
Ministry of Education and Scientific Research, Institute of Atomic Physics, Ministry of Research and Innovation and Institute of Atomic Physics and Universitatea Nationala de Stiinta si Tehnologie Politehnica Bucuresti, Romania;
Ministerstvo skolstva, vyskumu, vyvoja a mladeze SR, Slovakia;
National Research Foundation of South Africa, South Africa;
Swedish Research Council (VR) and Knut \& Alice Wallenberg Foundation (KAW), Sweden;
European Organization for Nuclear Research, Switzerland;
Suranaree University of Technology (SUT), National Science and Technology Development Agency (NSTDA) and National Science, Research and Innovation Fund (NSRF via PMU-B B05F650021), Thailand;
Turkish Energy, Nuclear and Mineral Research Agency (TENMAK), Turkey;
National Academy of  Sciences of Ukraine, Ukraine;
Science and Technology Facilities Council (STFC), United Kingdom;
National Science Foundation of the United States of America (NSF) and United States Department of Energy, Office of Nuclear Physics (DOE NP), United States of America.
In addition, individual groups or members have received support from:
FORTE project, reg.\ no.\ CZ.02.01.01/00/22\_008/0004632, Czech Republic, co-funded by the European Union, Czech Republic;
European Research Council (grant no. 950692), European Union;
Deutsche Forschungs Gemeinschaft (DFG, German Research Foundation) ``Neutrinos and Dark Matter in Astro- and Particle Physics'' (grant no. SFB 1258), Germany;
FAIR - Future Artificial Intelligence Research, funded by the NextGenerationEU program (Italy).

\end{acknowledgement}

%%%%%%%% Bibliography 
\bibliographystyle{utphys}   % Remember we use title in the biblio
\bibliography{bibliography}
%\input {bibliography.tex}  

%%%%%%%%%%%%%%%%%%%%%%%%%%%%%%%%
% Appendices: yours (if any) + authorlist
%%%%%%%%%%%%%%%%%%%%%%%%%%%%%%%%
\newpage
\appendix

%
%\input{} % put your appendices here (if any)
%

%%%%% Authorlist - please do not touch: handled by EB chairs 
\section{The ALICE Collaboration}
\label{app:collab}
% ALICE Collaboration author list for 2026-02-03
\begin{flushleft} 
\small

D.A.H.~Abdallah\,\orcidlink{0000-0003-4768-2718}\,$^{\rm 134}$, 
I.J.~Abualrob\,\orcidlink{0009-0005-3519-5631}\,$^{\rm 112}$, 
S.~Acharya\,\orcidlink{0000-0002-9213-5329}\,$^{\rm 49}$, 
K.~Agarwal\,\orcidlink{0000-0001-5781-3393}\,$^{\rm II,}$$^{\rm 23}$, 
G.~Aglieri Rinella\,\orcidlink{0000-0002-9611-3696}\,$^{\rm 32}$, 
L.~Aglietta\,\orcidlink{0009-0003-0763-6802}\,$^{\rm 24}$, 
N.~Agrawal\,\orcidlink{0000-0003-0348-9836}\,$^{\rm 25}$, 
Z.~Ahammed\,\orcidlink{0000-0001-5241-7412}\,$^{\rm 132}$, 
S.~Ahmad\,\orcidlink{0000-0003-0497-5705}\,$^{\rm 15}$, 
I.~Ahuja\,\orcidlink{0000-0002-4417-1392}\,$^{\rm 36}$
% ...

\section*{Affiliation Notes}

$^{\rm I}$ Deceased\\
$^{\rm II}$ Also at: INFN Trieste\\
$^{\rm III}$ Also at: Max-Planck-Institut fur Physik, Munich, Germany\\
$^{\rm IV}$ Also at: Czech Technical University in Prague (CZ)\\
$^{\rm V}$ Also at: Instituto de Fisica da Universidade de Sao Paulo\\
$^{\rm VI}$ Also at: Dipartimento DET del Politecnico di Torino, Turin, Italy\\
$^{\rm VII}$ Also at: Department of Applied Physics, Aligarh Muslim University, Aligarh, India\\
$^{\rm VIII}$ Also at: Institute of Theoretical Physics, University of Wroclaw, Poland\\
$^{\rm IX}$ Also at: Facultad de Ciencias, Universidad Nacional Aut\'{o}noma de M\'{e}xico, Mexico City, Mexico\\

\section*{Collaboration Institutes}

$^{1}$ A.I. Alikhanyan National Science Laboratory (Yerevan Physics Institute) Foundation, Yerevan, Armenia\\
$^{2}$ AGH University of Krakow, Cracow, Poland\\
$^{3}$ Bogolyubov Institute for Theoretical Physics, National Academy of Sciences of Ukraine, Kyiv, Ukraine\\
$^{4}$ Bose Institute, Department of Physics  and Centre for Astroparticle Physics and Space Science (CAPSS), Kolkata, India\\
$^{5}$ California Polytechnic State University, San Luis Obispo, California, United States\\
$^{6}$ Central China Normal University, Wuhan, China\\
$^{7}$ Centro de Aplicaciones Tecnol\'{o}gicas y Desarrollo Nuclear (CEADEN), Havana, Cuba\\
$^{8}$ Centro de Investigaci\'{o}n y de Estudios Avanzados (CINVESTAV), Mexico City and M\'{e}rida, Mexico\\
$^{9}$ Chicago State University, Chicago, Illinois, United States\\
$^{10}$ China Nuclear Data Center, China Institute of Atomic Energy, Beijing, China\\
$^{11}$ China University of Geosciences, Wuhan, China\\
$^{12}$ Chungbuk National University, Cheongju, Republic of Korea\\
$^{13}$ Comenius University Bratislava, Faculty of Mathematics, Physics and Informatics, Bratislava, Slovak Republic\\
$^{14}$ Creighton University, Omaha, Nebraska, United States\\
$^{15}$ Department of Physics, Aligarh Muslim University, Aligarh, India\\
$^{16}$ Department of Physics, Pusan National University, Pusan, Republic of Korea\\
$^{17}$ Department of Physics, Sejong University, Seoul, Republic of Korea\\
$^{18}$ Department of Physics, University of California, Berkeley, California, United States\\
$^{19}$ Department of Physics, University of Oslo, Oslo, Norway\\
$^{20}$ Department of Physics and Technology, University of Bergen, Bergen, Norway\\
$^{21}$ Dipartimento di Fisica, Universit\`{a} di Pavia, Pavia, Italy\\
$^{22}$ Dipartimento di Fisica dell'Universit\`{a} and Sezione INFN, Cagliari, Italy\\
$^{23}$ Dipartimento di Fisica dell'Universit\`{a} and Sezione INFN, Trieste, Italy\\
$^{24}$ Dipartimento di Fisica dell'Universit\`{a} and Sezione INFN, Turin, Italy\\
$^{25}$ Dipartimento di Fisica e Astronomia dell'Universit\`{a} and Sezione INFN, Bologna, Italy\\
$^{26}$ Dipartimento di Fisica e Astronomia dell'Universit\`{a} and Sezione INFN, Catania, Italy\\
$^{27}$ Dipartimento di Fisica e Astronomia dell'Universit\`{a} and Sezione INFN, Padova, Italy\\
$^{28}$ Dipartimento di Fisica `E.R.~Caianiello' dell'Universit\`{a} and Gruppo Collegato INFN, Salerno, Italy\\
$^{29}$ Dipartimento DISAT del Politecnico and Sezione INFN, Turin, Italy\\
$^{30}$ Dipartimento di Scienze MIFT, Universit\`{a} di Messina, Messina, Italy\\
$^{31}$ Dipartimento Interateneo di Fisica `M.~Merlin' and Sezione INFN, Bari, Italy\\
$^{32}$ European Organization for Nuclear Research (CERN), Geneva, Switzerland\\
$^{33}$ Faculty of Electrical Engineering, Mechanical Engineering and Naval Architecture, University of Split, Split, Croatia\\
$^{34}$ Faculty of Nuclear Sciences and Physical Engineering, Czech Technical University in Prague, Prague, Czech Republic\\
$^{35}$ Faculty of Physics, Sofia University, Sofia, Bulgaria\\
$^{36}$ Faculty of Science, P.J.~\v{S}af\'{a}rik University, Ko\v{s}ice, Slovak Republic\\
$^{37}$ Faculty of Technology, Environmental and Social Sciences, Bergen, Norway\\
$^{38}$ Frankfurt Institute for Advanced Studies, Johann Wolfgang Goethe-Universit\"{a}t Frankfurt, Frankfurt, Germany\\
$^{39}$ Fudan University, Shanghai, China\\
$^{40}$ Gauhati University, Department of Physics, Guwahati, India\\
$^{41}$ Helmholtz-Institut f\"{u}r Strahlen- und Kernphysik, Rheinische Friedrich-Wilhelms-Universit\"{a}t Bonn, Bonn, Germany\\
$^{42}$ Helsinki Institute of Physics (HIP), Helsinki, Finland\\
$^{43}$ High Energy Physics Group,  Universidad Aut\'{o}noma de Puebla, Puebla, Mexico\\
$^{44}$ Horia Hulubei National Institute of Physics and Nuclear Engineering, Bucharest, Romania\\
$^{45}$ HUN-REN Wigner Research Centre for Physics, Budapest, Hungary\\
$^{46}$ Indian Institute of Technology Bombay (IIT), Mumbai, India\\
$^{47}$ Indian Institute of Technology Indore, Indore, India\\
$^{48}$ INFN, Laboratori Nazionali di Frascati, Frascati, Italy\\
$^{49}$ INFN, Sezione di Bari, Bari, Italy\\
$^{50}$ INFN, Sezione di Bologna, Bologna, Italy\\
$^{51}$ INFN, Sezione di Cagliari, Cagliari, Italy\\
$^{52}$ INFN, Sezione di Catania, Catania, Italy\\
$^{53}$ INFN, Sezione di Padova, Padova, Italy\\
$^{54}$ INFN, Sezione di Pavia, Pavia, Italy\\
$^{55}$ INFN, Sezione di Torino, Turin, Italy\\
$^{56}$ INFN, Sezione di Trieste, Trieste, Italy\\
$^{57}$ Inha University, Incheon, Republic of Korea\\
$^{58}$ Institute for Gravitational and Subatomic Physics (GRASP), Utrecht University/Nikhef, Utrecht, Netherlands\\
$^{59}$ Institute of Experimental Physics, Slovak Academy of Sciences, Ko\v{s}ice, Slovak Republic\\
$^{60}$ Institute of Physics, Homi Bhabha National Institute, Bhubaneswar, India\\
$^{61}$ Institute of Physics of the Czech Academy of Sciences, Prague, Czech Republic\\
$^{62}$ Institute of Space Science (ISS), Bucharest, Romania\\
$^{63}$ Institut f\"{u}r Kernphysik, Johann Wolfgang Goethe-Universit\"{a}t Frankfurt, Frankfurt, Germany\\
$^{64}$ Instituto de Ciencias Nucleares, Universidad Nacional Aut\'{o}noma de M\'{e}xico, Mexico City, Mexico\\
$^{65}$ Instituto de F\'{i}sica, Universidade Federal do Rio Grande do Sul (UFRGS), Porto Alegre, Brazil\\
$^{66}$ Instituto de F\'{\i}sica, Universidad Nacional Aut\'{o}noma de M\'{e}xico, Mexico City, Mexico\\
$^{67}$ iThemba LABS, National Research Foundation, Somerset West, South Africa\\
$^{68}$ Jeonbuk National University, Jeonju, Republic of Korea\\
$^{69}$ Korea Institute of Science and Technology Information, Daejeon, Republic of Korea\\
$^{70}$ Laboratoire de Physique Subatomique et de Cosmologie, Universit\'{e} Grenoble-Alpes, CNRS-IN2P3, Grenoble, France\\
$^{71}$ Lawrence Berkeley National Laboratory, Berkeley, California, United States\\
$^{72}$ Lund University Department of Physics, Division of Particle Physics, Lund, Sweden\\
$^{73}$ Marietta Blau Institute, Vienna, Austria\\
$^{74}$ Nagasaki Institute of Applied Science, Nagasaki, Japan\\
$^{75}$ Nara Women{'}s University (NWU), Nara, Japan\\
$^{76}$ National Centre for Nuclear Research, Warsaw, Poland\\
$^{77}$ National Institute of Science Education and Research, Homi Bhabha National Institute, Jatni, India\\
$^{78}$ National Nuclear Research Center, Baku, Azerbaijan\\
$^{79}$ National Research and Innovation Agency - BRIN, Jakarta, Indonesia\\
$^{80}$ Niels Bohr Institute, University of Copenhagen, Copenhagen, Denmark\\
$^{81}$ Nikhef, National institute for subatomic physics, Amsterdam, Netherlands\\
$^{82}$ Nuclear Physics Group, STFC Daresbury Laboratory, Daresbury, United Kingdom\\
$^{83}$ Nuclear Physics Institute of the Czech Academy of Sciences, Husinec-\v{R}e\v{z}, Czech Republic\\
$^{84}$ Oak Ridge National Laboratory, Oak Ridge, Tennessee, United States\\
$^{85}$ Ohio State University, Columbus, Ohio, United States\\
$^{86}$ Physics department, Faculty of science, University of Zagreb, Zagreb, Croatia\\
$^{87}$ Physics Department, Panjab University, Chandigarh, India\\
$^{88}$ Physics Department, University of Jammu, Jammu, India\\
$^{89}$ Physics Program and International Institute for Sustainability with Knotted Chiral Meta Matter (WPI-SKCM$^{2}$), Hiroshima University, Hiroshima, Japan\\
$^{90}$ Physikalisches Institut, Eberhard-Karls-Universit\"{a}t T\"{u}bingen, T\"{u}bingen, Germany\\
$^{91}$ Physikalisches Institut, Ruprecht-Karls-Universit\"{a}t Heidelberg, Heidelberg, Germany\\
$^{92}$ Physik Department, Technische Universit\"{a}t M\"{u}nchen, Munich, Germany\\
$^{93}$ Politecnico di Bari and Sezione INFN, Bari, Italy\\
$^{94}$ Research Division and ExtreMe Matter Institute EMMI, GSI Helmholtzzentrum f\"ur Schwerionenforschung GmbH, Darmstadt, Germany\\
$^{95}$ Saga University, Saga, Japan\\
$^{96}$ Saha Institute of Nuclear Physics, Homi Bhabha National Institute, Kolkata, India\\
$^{97}$ School of Physics and Astronomy, University of Birmingham, Birmingham, United Kingdom\\
$^{98}$ Secci\'{o}n F\'{\i}sica, Departamento de Ciencias, Pontificia Universidad Cat\'{o}lica del Per\'{u}, Lima, Peru\\
$^{99}$ SUBATECH, IMT Atlantique, Nantes Universit\'{e}, CNRS-IN2P3, Nantes, France\\
$^{100}$ Sungkyunkwan University, Suwon City, Republic of Korea\\
$^{101}$ Suranaree University of Technology, Nakhon Ratchasima, Thailand\\
$^{102}$ Technical University of Ko\v{s}ice, Ko\v{s}ice, Slovak Republic\\
$^{103}$ The Henryk Niewodniczanski Institute of Nuclear Physics, Polish Academy of Sciences, Cracow, Poland\\
$^{104}$ The University of Texas at Austin, Austin, Texas, United States\\
$^{105}$ Universidad Aut\'{o}noma de Sinaloa, Culiac\'{a}n, Mexico\\
$^{106}$ Universidade de S\~{a}o Paulo (USP), S\~{a}o Paulo, Brazil\\
$^{107}$ Universidade Estadual de Campinas (UNICAMP), Campinas, Brazil\\
$^{108}$ Universidade Federal do ABC, Santo Andre, Brazil\\
$^{109}$ Universitatea Nationala de Stiinta si Tehnologie Politehnica Bucuresti, Bucharest, Romania\\
$^{110}$ University of Cape Town, Cape Town, South Africa\\
$^{111}$ University of Derby, Derby, United Kingdom\\
$^{112}$ University of Houston, Houston, Texas, United States\\
$^{113}$ University of Jyv\"{a}skyl\"{a}, Jyv\"{a}skyl\"{a}, Finland\\
$^{114}$ University of Kansas, Lawrence, Kansas, United States\\
$^{115}$ University of Liverpool, Liverpool, United Kingdom\\
$^{116}$ University of Science and Technology of China, Hefei, China\\
$^{117}$ University of Silesia in Katowice, Katowice, Poland\\
$^{118}$ University of South-Eastern Norway, Kongsberg, Norway\\
$^{119}$ University of Tennessee, Knoxville, Tennessee, United States\\
$^{120}$ University of the Witwatersrand, Johannesburg, South Africa\\
$^{121}$ University of Tokyo, Tokyo, Japan\\
$^{122}$ University of Tsukuba, Tsukuba, Japan\\
$^{123}$ Universit\"{a}t M\"{u}nster, Institut f\"{u}r Kernphysik, M\"{u}nster, Germany\\
$^{124}$ Universit\'{e} Clermont Auvergne, CNRS/IN2P3, LPC, Clermont-Ferrand, France\\
$^{125}$ Universit\'{e} de Lyon, CNRS/IN2P3, Institut de Physique des 2 Infinis de Lyon, Lyon, France\\
$^{126}$ Universit\'{e} de Strasbourg, CNRS, IPHC UMR 7178, F-67000 Strasbourg, France, Strasbourg, France\\
$^{127}$ Universit\'{e} Paris-Saclay, Centre d'Etudes de Saclay (CEA), IRFU, D\'{e}partment de Physique Nucl\'{e}aire (DPhN), Saclay, France\\
$^{128}$ Universit\'{e}  Paris-Saclay, CNRS/IN2P3, IJCLab, Orsay, France\\
$^{129}$ Universit\`{a} degli Studi di Foggia, Foggia, Italy\\
$^{130}$ Universit\`{a} del Piemonte Orientale, Vercelli, Italy\\
$^{131}$ Universit\`{a} di Brescia, Brescia, Italy\\
$^{132}$ Variable Energy Cyclotron Centre, Homi Bhabha National Institute, Kolkata, India\\
$^{133}$ Warsaw University of Technology, Warsaw, Poland\\
$^{134}$ Wayne State University, Detroit, Michigan, United States\\
$^{135}$ Yale University, New Haven, Connecticut, United States\\
$^{136}$ Yildiz Technical University, Istanbul, Turkey\\
$^{137}$ Yonsei University, Seoul, Republic of Korea\\
$^{138}$ Affiliated with an institute formerly covered by a cooperation agreement with CERN\\
$^{139}$ Affiliated with an international laboratory covered by a cooperation agreement with CERN.\\

\end{flushleft} 
  
\end{document}